\documentclass{article}
\usepackage[a4paper, total={6in, 8in}]{geometry}
\usepackage{authblk}
\usepackage{amsmath}
\usepackage{amssymb}
\usepackage{bm}
\usepackage{listings}
\usepackage{jlcode}
\numberwithin{equation}{section}
\usepackage{stackengine} % for decent underbars
\usepackage{subfig}
\usepackage{multirow}
\usepackage{booktabs}
\usepackage{algorithm} % floating algorithm environment with algorithmic keywords
\usepackage{algpseudocode} % from algorithmicx
\algrenewcommand\algorithmicindent{1.1em}
\usepackage{mathtools}
\usepackage{soul}
\usepackage{hyperref}
\usepackage{url}

\newcommand{\etal}{\textit{et al}.}
\newcommand{\ie}{\textit{i}.\textit{e}.}
\newcommand{\eg}{\textit{e}.\textit{g}.}
\newcommand\ubar[1]{\stackunder[1.2pt]{\(#1\)}{\rule{.8ex}{.075ex}}}

\newcommand{\keywords}[1]
{
  \small	
  \textbf{\textit{Keywords---}} #1
}

\begin{document}

\title{Extending JumpProcess.jl for fast point process simulation with time-varying intensities}

\author[1]{Guilherme Augusto Zagatti}
\author[3]{Samuel A. Isaacson}
\author[4]{Christopher Rackauckas}
\author[5]{Vasily Ilin}
\author[1, 2]{See-Kiong Ng}
\author[1, 2]{Stéphane Bressan}
\affil[1]{Institute of Data Science, National University of Singapore, Singapore}
\affil[2]{School of Computing, National University of Singapore, Singapore}
\affil[3]{Department of Mathematics and Statistics, Boston University}
\affil[4]{Computer Science and AI Laboratory (CSAIL), Massachusetts Institute of Technology}
\affil[5]{Department of Mathematics, University of Washington}

\hypersetup{
  pdftitle = {Extending JumpProcess.jl for fast point process simulation with time-varying intensities},
  pdfsubject = {JuliaCon 2023 Proceedings},
  pdfauthor = {Guilherme Augusto Zagatti, Samuel A. Isaacson, Christopher Rackauckas, Vasily Ilin, See-Kiong Ng, Stéphane Bressan},
  pdfkeywords = {Julia, Simulation, Jump process, Point process},
}

\maketitle

\abstract{Point processes model the occurrence of a countable number of random points over some support. They can model diverse phenomena, such as chemical reactions, stock market transactions and social interactions. We show that \texttt{JumpProcesses.jl} is a fast, general-purpose library for simulating point processes. \texttt{JumpProcesses.jl} was first developed for simulating jump processes via stochastic simulation algorithms (SSAs) (including Doob's method, Gillespie's methods, and Kinetic Monte Carlo methods). Historically, jump processes have been developed in the context of dynamical systems to describe dynamics with discrete jumps. In contrast, the development of point processes has been more focused on describing the occurrence of random events. In this paper, we bridge the gap between the treatment of point and jump process simulation. The algorithms previously included in \texttt{JumpProcesses.jl} can be mapped to three general methods developed in statistics for simulating evolutionary point processes. Our comparative exercise revealed that the library initially lacked an efficient algorithm for simulating processes with variable intensity rates. We, therefore, extended \texttt{JumpProcesses.jl} with a new simulation algorithm, \texttt{Coevolve}, that enables the rapid simulation of processes with locally-bounded variable intensity rates. It is now possible to efficiently simulate any point process on the real line with a non-negative, left-continuous, history-adapted and locally bounded intensity rate coupled or not with differential equations. This extension significantly improves the computational performance of \texttt{JumpProcesses.jl} when simulating such processes, enabling it to become one of the few readily available, fast, general-purpose libraries for simulating evolutionary point processes.}

\section{Introduction}

Methods for simulating the trajectory of evolutionary point processes can be split into exact and inexact methods. Exact methods describe the realization of each point in the process chronologically. This exactness avoids bias from numerical approximations, but such methods can suffer from reduced performance when simulating systems with large populations (where numerous events can fire within a short period since every single point needs to be accounted for). Inexact methods trade accuracy for speed by simulating the total number of events in successive intervals. They are popular in biochemical applications, e.g. \( \tau \)-leap methods~\cite{gillespie2001}, which often require the simulation of chemical reactions in systems with large molecular populations.

Previously, point process simulation library development focused primarily on univariate processes with exotic intensities, or large systems with conditionally constant intensities, but not on both. As such, there was no widely used general-purpose software for efficiently simulating compound point processes in large systems with time-dependent rates. To enable the efficient simulation of such processes, we contribute the \texttt{Coevolve} aggregator to \texttt{JumpProcesses.jl}, a core component of the popular \texttt{DifferentialEquations.jl} library~\cite{rackauckas2017}. The implemented algorithm improves the COEVOLVE algorithm described in~\cite{farajtabar2017} from where it borrows its name. Among other improvements, our algorithm supports any process with locally bounded conditional intensity rates, adapts to intensity rates that can change between jumps, can be coupled with differential equations, and avoids both the unnecessary re-computation of randomly generated numbers and the computation of the intensity rate when its lower bound is available. This extension of \texttt{JumpProcesses.jl} dramatically boosts the computational performance of the library in simulating processes with intensities that have an explicit dependence on time and/or other continuous variables,  significantly expanding the type of models that can be efficiently simulated. Widely-used point processes with such intensities include compound inhomogeneous Poisson, Hawkes, and stress-release processes --- all described in~\cite{daley2003}. Since \texttt{JumpProcesses.jl} is a member of Julia's SciML organization, it also becomes easier, and more feasible, to incorporate compound point processes with explicit time-dependent rates into a wide variety of applications and higher-level analyses. With our new additions we bump \texttt{JumpProcesses.jl} to version 9.7\footnote{All examples and benchmarks in this paper use this version of the library}.

In this paper, we bridge the gap between simulation methods developed in statistics and biochemistry, which led us to the development of \texttt{Coevolve}. First, we briefly introduce evolutionary point processes. Next, since all simulation methods require a basic understanding of simulation methods for the Poisson homogeneous process, we first describe such methods. Then, we identify and discuss three general, exact methods. In the second part of this paper, we describe the algorithms in \texttt{JumpProcesses.jl} and how they relate to the literature. We highlight our contribution \texttt{Coevolve}, investigate the correctness of our implementation and provide performance benchmarks to demonstrate its value. The paper concludes by discussing potential improvements.

\section{The evolutionary point process} \label{sec:notation}

The evolutionary point process is a stochastic collection of marked points over a one-dimensional support. They are exhaustively described in~\cite{daley2003}. The likelihood of any evolutionary point process is fully characterized by its conditional intensity,
\begin{equation}\label{eq:lambda}
  \lambda^\ast (t) \equiv \lambda(t \mid H_{t^-} ) = \frac{p^\ast(t)}{1 - \int_{t^-}^{t_n} p^\ast(u) \, du},
\end{equation}
and conditional mark distribution, \( f^*(k | t) \) --- see~Chapter 7~\cite{daley2003}. Here \( H_{t^-} = \{ (t_n, k_n) \mid 0 \leq t_n \leq t \} \) denotes the internal history of the process up to but not including \( t \), the superscript \( \ast \) denotes the conditioning of any function on \( H_{t^-} \), and \( p^\ast(t) \) is the density function corresponding to the probability of an event taking place at time \( t \) given \( H_{t^-} \). We can interpret the conditional intensity as the likelihood of observing a point in the next infinitesimal unit of time, given that no point has occurred since the last observed point in \( H_{t^-} \). Lastly, the mark distribution denotes the density function corresponding to the probability of observing mark \( k \) given the occurrence of an event at time \( t \) and internal history \( H_{t^-} \).

\section{The homogeneous process} \label{sec:method-poisson}

A homogeneous process can be simulated using properties of the Poisson process, which allow us to describe two equivalent sampling procedures. The first procedure consists of drawing successive inter-arrival times. The distance between any two points in a homogeneous process is distributed according to the exponential distribution --- see Theorem 7.2~\cite{last2017}. Given the homogeneous process with intensity $\lambda$, then the distance \( \Delta t \) between two points is distributed according to $\Delta t \sim \exp(\lambda)$. Draws from the exponential distribution can be performed by drawing from a uniform distribution in the interval $[0, 1]$. If $V \sim U[0, 1]$, then \( T = - \ln(V) / \lambda \sim \exp(1) \). (Note, however, in Julia the optimized Ziggurat-based method used in the \texttt{randexp} stdlib function is generally faster than this \textit{inverse} method for sampling a unit exponential random variable.) When a point process is homogeneous, the \textit{inverse} method of Subsection~\ref{subsec:sim-inverse} reduces to this approach. Thus, we defer the presentation of this Algorithm to the next section.

The second procedure uses the fact that Poisson processes can be represented as a mixed binomial process with a Poisson mixing distribution --- see Proposition 3.5~\cite{last2017}. In particular, the total number of points of a Poisson homogeneous process in \( [0, T) \) is distributed according to \( \mathcal{N} (T) \sim \operatorname{Poisson}( \lambda T ) \) and the location of each point within the region is distributed according to the uniform distribution \( t_n \sim U[0, T] \).

% Tau leaping methods in Subsection~\ref{subsec:sim-tau} use a similar procedure to focus only on sampling the total number of points in successive intervals.

% Algorithm~\ref{algo:sim-mixed} lists the implementation of this approach.

% \begin{algorithm}[h]
% \begin{algorithmic}[1]
%   \Procedure{MixedBinomialMethod}{\( [0, T) \), \( \lambda \),}
%   \State initialize the history \( H_{T^-} \leftarrow \{ \} \)
%   \State draw the total number of points \( N \sim \operatorname{Poisson}( \lambda T ) \)
%   \For{\(n \leftarrow 1, N\)}
%       \State draw the location \( t_n \sim U[0, T] \)
%       \State update the history \( H_{T^-} \leftarrow H_{T^-} \cup (t_n) \)
%   \EndFor
%   \State \Return \( H_{T^-} \)
%   \EndProcedure
% \end{algorithmic}
% \caption{The mixed binomial method for simulating a homogeneous point process over a fixed duration of time \( [0, T) \).}
% \label{algo:sim-mixed}
% \end{algorithm}

\section{Exact simulation methods} \label{sec:method-exact}

\subsection{Inverse methods} \label{subsec:sim-inverse}

The \textit{inverse} method leverages Theorem 7.4.I~\cite{daley2003} which states that every simple point process\footnote{A simple point process is a process in which the probability of observing more than one point in the same location is zero.} can be transformed to a homogeneous Poisson process with unit rate via the compensator. Let \( t_n \) be the time in which the \( n \)-th chronologically sorted event took place and \( t_0 \equiv 0 \), we define the compensator as:
\begin{equation} \label{eqn:compensator}
  \Lambda^\ast (t_n) \equiv \tilde{t}_n \equiv \int_0^{t_n} \lambda^\ast (u) du
\end{equation}
The transformed data \( \tilde{t}_n \) forms a homogeneous Poisson process with unit rate. Now, if this is the case, then the transformed interval is distributed according to the exponential distribution.
\begin{equation}\label{eqn:inverse}
  \Delta \tilde{t}_n \equiv \tilde{t}_n - \tilde{t}_{n-1} \sim \exp(1)
\end{equation}
The idea is to draw realizations from the unit rate Exponential process and solve Equation~\ref{eqn:inverse} for \( t_n \)  to determine the next event/firing time. We illustrate this in Algorithm~\ref{algo:sim-inverse} where we adapt Algorithm 7.4~\cite{daley2003}.

Whenever the conditional intensity is constant between two points, Equation~\ref{eqn:inverse} can be solved analytically. Let \( \lambda^\ast \, (t) = \lambda_{n-1} , \forall t_{n-1} \leq t < t_n \), then
\begin{equation}
\begin{split}
  &\int_{t_{n-1}}^{t_n} \lambda^\ast \, (u) \, du = \Delta \tilde{t}_{n} \iff \\
  &\lambda_{n-1} (t_n - t_{n-1}) = \Delta \tilde{t}_n \iff \\
  &t_n = t_{n-1} + \frac{\Delta \tilde{t}_n}{\lambda_{n-1}}.
\end{split}
\end{equation}
Which is equivalent to drawing the next realization time from the re-scaled exponential distribution \( \Delta t_n \sim \exp(\lambda_{n-1}) \). As we will see in Subsection~\ref{algo:sim-thinning}, this implies that the \textit{inverse} and \textit{thinning} methods are the same whenever the conditional intensity is constant between jumps.

% see here on computing the inverse of integrals:
% https://math.stackexchange.com/questions/1467784/inverse-of-a-functions-integral
The main drawback of the \textit{inverse} method is that the root finding problem defined in Equation~\ref{eqn:inverse} often requires a numerical solution. To get around a similar obstacle in the context of the piecewise deterministic Markov process, Veltz~\cite{veltz2015} proposes a change of variables in time that recasts the root finding problem into an initial value problem. He denotes his method \textit{CHV}.

Piecewise deterministic Markov processes are composed of two parts: the jump process and the piecewise ODE that changes stochastically at jump times --- see Lemaire~\etal~\cite{lemaire2018} for a formal definition. Therefore, it is easy to employ \textit{CHV} in our case by setting the ODE part to zero throughout time. Adapting from Veltz~\cite{veltz2015}, we can determine the model jump time \( t_n \) after sampling \( \Delta \tilde{t}_n \sim \exp(1) \) by solving the following initial value problem until \( \Delta \tilde{t}_n \).
\begin{equation} \label{eqn:chv-simple}
    t (0) = t_{n-1} \text{ , } \frac{d t}{d \tilde{t} } = \frac{1}{\lambda^\ast (t)}
\end{equation}
Looking back at Equation~\ref{eqn:compensator}, we note that it is a one-to-one mapping between \( t \) and \( \tilde{t} \) which makes it completely natural to write \( t(\Delta \tilde{t}_n) \equiv \Lambda^{\ast-1} (\tilde{t}_{n-1} + \Delta \tilde{t}_n) \).

Alternatively, when the intensity function is differentiable between jumps we can go even further by recasting the jump problem as a piecewise deterministic Markov process. Let \( \lambda^\ast_n \equiv \lambda^\ast(t_n) \), then the flow \( \varphi_{t-t_n}( \lambda^\ast_n ) \) maps the initial value of the conditional intensity at time \( t_n \) to its value at time \( t \). In other words, the flow describes the deterministic evolution of the conditional intensity function over time. Next, denote \( \mathbf{1}( \cdot ) \) as the indicator function, then the conditional intensity function can be re-written as a jump process:
\begin{equation} \label{eqn:conditional-jump}
  \lambda^\ast (t) = \sum_{n \geq 1} \varphi_{t - t_{n-1}} ( \lambda_{n-1} ) \mathbf{1}(t_{n-1} \leq t < t_n).
\end{equation}
According to Meiss~\cite{meiss2017}, if \( \varphi_t ( \cdot ) \) is a flow, then it is a solution to the initial value problem:
\begin{equation}
  \varphi_{0} (\lambda_n^\ast) = \lambda_n^\ast \text{ , }
  \frac{d}{dt} \varphi_{t-t_n} (\lambda_n^\ast) = g(\varphi_{t-t_n} (\lambda_n^\ast))
\end{equation}
where \( g: \mathbb{R}^+ \to \mathbb{R} \) is the vector field of \( \lambda^\ast \) such that \( d \lambda^\ast / dt = g( \lambda^\ast ) \).

Based on Equation~\ref{eq:lambda}, we find that the probability of observing an interval longer than \( s \) given internal history \( H_{t^-} \) is equivalent to:
\begin{equation} \label{eqn:transition-rate}
\begin{split}
  &\Pr(t_n - t_{n-1} > s \mid H_{t^-}) = 1 - \int_{t_{n-1}}^{t_{n-1} + s} p^\ast(u) du = \\
    &=\exp \left( -\int_{t_{n-1}}^{t_{n-1} + s} \lambda^\ast (u) du \right) = \\
    &=\exp \left( -\int_{t_{n-1}}^{t_{n-1} + s} \varphi_{u-t_{n-1}} (\lambda_{n-1}^\ast) du \right)
\end{split}
\end{equation}

Equations~\ref{eqn:conditional-jump} and~\ref{eqn:transition-rate} define a piecewise deterministic Markov process satisfying the conditions of Theorem 3.1~\cite{veltz2015}. In this case, we find \( t_n \) by solving the following initial value problem from \( 0 \) to \( \Delta \tilde{t}_n \sim \exp(1) \).
\begin{equation} \label{eqn:chv-full}
  \begin{cases}
  \begin{aligned}
    & \lambda^\ast (t(0)) = \lambda^\ast (t_{n-1}) \text{ , } \frac{d \lambda^\ast}{d \tilde{t}} =  \frac{g(\lambda^\ast(t))}{\lambda^\ast (t)} \\
    & t (0) = t_{n-1} \text{ , } \frac{d t}{d \tilde{t} } = \frac{1}{\lambda^\ast (t)}.
  \end{aligned}
  \end{cases}
\end{equation}
This problem specifies how the conditional intensity and model time evolve with respect to the transformed time. The solution to Equation~\ref{eqn:inverse} is then given by \( ( t_n = t(\Delta \tilde{t}_n), \lambda^\ast(t(\Delta \tilde{t}_n)) =  \lambda^\ast(t_n)) \). 

In Algorithm~\ref{algo:sim-inverse}, we can implement the CHV method by solving either Equation~\ref{eqn:chv-simple} or Equation~\ref{eqn:chv-full} instead of Equation~\ref{eqn:inverse}. We denote the first specification as \textit{CHV simple} and the second as \textit{CHV full}. Note that \textit{CHV full} requires that the conditional intensity be piecewise differentiable. The algorithmic complexity is then determined by the ODE solver and no root-finding is required. In Section~\ref{subsec:benchmark}, we will show that there are substantial differences in performance between them with the full specification being faster. %For more discussions on the links between point processes and Markov processes see Chapter 10~\cite{daley2007}.

Another concern with Algorithm~\ref{algo:sim-inverse} is updating and drawing from the conditional mark distribution in Line~\ref{line:inverse-mark-sample}, and updating the conditional intensity in Line~\ref{line:inverse-history-update}. Assume a process with \( K \) number of marks. A naive implementation of Line~\ref{line:inverse-history-update} scales with the number of marks as \( O(K) \) since \( \lambda^\ast \) is usually constructed as the sum of \( K \) independent processes, each of which requires updating the conditional intensity rate. Likewise, drawing from the mark distribution in Line~\ref{line:inverse-mark-sample} usually involves drawing from a categorical distribution whose naive implementations also scales with the number of marks as \( O(K) \).

Finally, Algorithm~\ref{algo:sim-inverse} is not guaranteed to terminate in finite time since one might need to sample many points before \( t_n > T \). The sampling rate can be especially high when simulating the process in a large population with self-exciting encounters. In biochemistry, Salis and Kaznessis~\cite{salis2005} partition a large system of chemical reactions into two: fast and slow reactions. While they approximate the fast reactions with a Gaussian process, the slow reactions are solved using a variation of the inverse method. They obtain an equivalent expression for the rate of slow reactions as in Equation~\ref{eqn:inverse}, which is integrated with the Euler method.

\begin{algorithm}[h]
\begin{algorithmic}[1]
  \Procedure{InverseMethod}{\( [0, T) \), \( \lambda^\ast \), \( f^\ast \),}
    \State initialize the history \( H_{T^-} \leftarrow \{ \} \)
    \State set \( n \leftarrow 0, t \leftarrow  0 \)
    \While{\( t < T \)}
      \State \( n \leftarrow n + 1 \)
      \State draw \( \Delta \tilde{t}_n \sim \exp(1) \)
      \State find the next event time \( t_n \) by solving Equation~\ref{eqn:inverse} or~\ref{eqn:chv-full}
      \State update \( f^\ast \) and draw the mark \( k_n \sim f^\ast \, (k \mid t_n) \) \label{line:inverse-mark-sample}
      \State update the history \( H_{T^-} \leftarrow H_{T^-} \cup (t_n, k_n) \) and \( \lambda^\ast \) \label{line:inverse-history-update}
    \EndWhile
    \State \Return \( H_{T^-} \)
  \EndProcedure
\end{algorithmic}
\caption{The \textit{inverse} method for simulating a marked evolutionary point process over a fixed duration of time \( [0, T) \).}
\label{algo:sim-inverse}
\end{algorithm}

\subsection{Thinning methods} \label{subsec:sim-thinning}

\textit{Thinning} methods are one of the most popular methods for simulating point processes. The main idea is to successively sample a homogeneous process, then thin the obtained points with the conditional intensity of the original process. As stated in Proposition 7.5.I~\cite{daley2003}, this procedure simulates the target process by construction. The advantage of \textit{thinning} over \textit{inverse} methods is that the former only requires the evaluation of the conditional intensity function while the latter requires computing the inverse of its integrated form~\cite{daley2003}.

\textit{Thinning} algorithms have been proposed in different forms~\cite{daley2003}. The Shedler-Lewis algorithm can simulate processes with bounded intensity~\cite{lewis1976}. The classical algorithm from Ogata~\cite{ogata1981} overcomes this limitation and only requires the local boundedness of the conditional intensity. The advantage of Ogata's algorithm and its variations is that it can simulate processes with potentially unbounded intensity, such as self-exciting ones. As long as the intensity conditioned on the simulated history remains locally bounded, it is possible to simulate subsequent points indefinitely.

In biochemistry, the \textit{thinning} method was popularized by Gillespie~\cite{gillespie1976,gillespie1977}. For this reason, this method is also called the \textit{Gillespie} method. Gillespie himself called it the \textit{direct} method or the \textit{stochastic simulation algorithm}. Gillespie introduced the \textit{thinning} method in the context of simulating chemical reactions of well-stirred systems. He developed a stochastic model for molecule interactions from physics principles without any references to the point process theory developed in this section. His model of chemical interactions is equivalent to a marked Poisson process with constant conditional intensity between jumps. The model consists of distinct populations of molecular species that interact through several reaction channels. A chemical reaction consists of a Poisson process that transforms a set of molecules of some type into a set of molecules of another type. What Gillespie calls the master equation can be deduced from the \textit{superposition theorem} --- Theorem 3.3~\cite{last2017}.

Alternatively, in biochemistry, \textit{thinning} methods are known as \textit{rejection} algorithms. Than~\etal~\cite{thanh2014,thanh2017} proposed the \textit{rejection-based algorithm with composition-rejection search}, yet another more sophisticated variation of the \textit{thinning} method. In this case, the procedure groups similar processes together. For each group, an upper- and lower-bound conditional intensity is used for thinning. A similar procedure is also described in \cite{slepoy2008}, in which the authors refer to their algorithm as \textit{kinetic Monte Carlo}.

In Algorithm~\ref{algo:sim-thinning}, we modify Algorithm 7.5.IV~\cite{daley2003} to incorporate the idea of a lower bound for the conditional intensity from~\cite{thanh2017}. To implement the algorithm, we define three functions, \( \bar{B}^\ast (t) = \bar{B}(t \mid H_t) \), \( \ubar{B}^\ast (t) = \ubar{B}(t\mid H_t) \) and \( L^\ast(t) = L(t \mid H_t) \), that characterize the local boundedness condition such that:
\begin{equation} \label{eq:thinning-condition}
\begin{multlined}
  \lambda^\ast \, (t + u)  \leq \bar{B}^\ast(t)  \text{ and } \lambda^\ast \, (t + u)  \geq \ubar{B}^\ast(t), \\ \forall \, 0 \leq u \leq L^\ast(t).
\end{multlined}
\end{equation}
The tighter the bound on \( \bar{B}^\ast(t) \), the lower the number of samples discarded. Since looser bounds lead to less efficient algorithms, the art, when simulating via \textit{thinning}, is to find the optimal balance between the local supremum of the conditional intensity \( \bar{B}^\ast(t) \) and the duration of the local interval \( L^\ast(t) \). On the other hand, the infimum \( \ubar{B}^\ast(t) \) can be used to avoid the evaluation of \( \lambda^\ast \, (t + u) \) in Line~\ref{line:short-circuit} of Algorithm~\ref{algo:next-time-thinning} which often can be expensive.

When the conditional intensity is constant between jumps such that \( \lambda^\ast \, (t) = \lambda_{n-1} , \forall t_{n-1} \leq t < t_n \), let \( \bar{B}^\ast(t) = \ubar{B}^\ast(t) = \lambda_{n-1} \) and \( L^\ast(t) = \infty \). We have that for any \( u \sim \exp(1 \; / \; \bar{B}^\ast(t)) =  \exp(\lambda_{n-1})\) and \( v \sim U[0, 1] \), \( u < L^\ast(t) = \infty \) and \( v < \lambda^\ast \, (t + u) \; / \; \bar{B}^\ast(t) = 1 \). Therefore, we advance the internal history for every iteration of Algorithm~\ref{algo:sim-thinning}. In this case, the bound \( \bar{B}^\ast(t) \) is as tight as possible, and this method becomes the same as the \textit{inverse} method of Subsection~\ref{subsec:sim-inverse}.

We can draw many more connections between the \textit{thinning} and \textit{inverse} methods. Lemaire~\etal~\cite{lemaire2018} propose a version of the \textit{thinning} algorithm for Piecewise Deterministic Markov Processes which does not use the local interval \( L^\ast \) for rejection --- this is equivalent to \( L^\ast(t) = \infty \) ---, and does not assume the upper bound \( \bar{B}^\ast(t) \) is constant over \( L^\ast(t) \). The efficiency of their algorithm depends on the assumption that the stochastic process determined by \( \bar{B}^\ast(t) \) can be efficiently inverted such that candidate times can be efficiently obtained using Equation~\ref{eqn:compensator}. They propose an optimal bound as a piecewise constant function partitioned in such a way that it envelopes the intensity function as strictly as possible. They then show that under certain conditions the stochastic process determined by \( \bar{B}^\ast(t) \) converges in distribution to the target conditional intensity as the partitions of the optimal boundary converge to zero. Although their simulation approach does not exactly match ours, it suggests some properties between the \textit{thinning} and the \textit{inverse} method that we could investigate in the future. Among other things, the efficiency of \textit{thinning} compared to \textit{inversion} most likely depends on the rejection rate obtained by the former and the number of steps required by the ODE solver for the latter.

While \textit{thinning} algorithms avoid the issue of directly computing the inverse of the integrated conditional intensity, they increase the number of time steps needed in the sampling algorithm as we are now sampling from a process with an increased intensity relative to the original process. Moreover, like the \textit{inverse} method, \textit{thinning} algorithms can also face issues related with drawing from the conditional mark distribution --- Line~\ref{line:thinning-mark-sample} of Algorithm~\ref{algo:sim-thinning} ---, and updating the conditional intensity --- Line~\ref{line:lambda-update} of Algorithm~\ref{algo:next-time-thinning} --- and the mark distribution --- Line~\ref{line:thinning-history-update} of Algorithm~\ref{algo:sim-thinning}.

\begin{algorithm}[h]
\begin{algorithmic}[1]
  \Procedure{ThinningMethod}{\( [0, T) \), \( \lambda^\ast \), \( f^\ast \),}
    \State initialize the history \( H_{T^-} \leftarrow \{ \} \)
    \State set \( n \leftarrow 0, t \leftarrow 0 \)
    \While{true}
      \State \( t \leftarrow \operatorname{TimeViaThinning}([t, T), H_{T^-}, \lambda^\ast) \)
      \If{\(t \geq T \)}
        \State \textbf{break}
      \EndIf
      \State \( n \leftarrow n + 1 \)
      \State \( t_n \leftarrow t \)
      \State update  \( f^\ast \) and draw the mark \( k_n \sim f^\ast \, (k \mid t_n) \) \label{line:thinning-mark-sample}
      \State update the history \( H_{T^-} \leftarrow H_{T^-} \cup (t_n, k_n) \) \label{line:thinning-history-update}
    \EndWhile
    \State \Return \( H_{T^-} \)
  \EndProcedure
\end{algorithmic}
\caption{The \textit{thinning} method for simulating a marked evolutionary point process over a fixed duration of time \( [0, T) \).}
\label{algo:sim-thinning}
\end{algorithm}

\begin{algorithm}[h]
\begin{algorithmic}[1]
  \Procedure{TimeViaThinning}{\([t, T) \), \( \lambda^\ast \), \( H_{t} \),}
      \While{\( t < T \)}
        \State update \( \lambda^\ast \) \label{line:lambda-update}
        \State find \( \bar{B}^\ast (t) \), \( \ubar{B}^\ast (t) \) and \( L^\ast(t) \) which satisfy Eq.~\ref{eq:thinning-condition}
        \State draw \( u \sim \exp(\bar{B}^\ast(t)) \) and \( v \sim U[0, \bar{B}^\ast(t)] \) \label{line:short-circuit}
        \If{\( u > L^\ast(t) \)}
          \State \( t \leftarrow t + L^\ast(t) \)
          \State \textbf{next}
        \EndIf
        \If{\( ( v > \ubar{B}^\ast(t) ) \) and \( ( v > \lambda^\ast \, (t + u) ) \)}
          \State \( t \leftarrow t + u \)
          \State \textbf{next}
        \EndIf
        \State \( t \leftarrow t + u \)
        \State \textbf{break}
      \EndWhile
      \State \Return t
  \EndProcedure
\end{algorithmic}
\caption{Generates the next event time via \textit{thinning}.}
\label{algo:next-time-thinning}
\end{algorithm}

\subsection{Queuing methods} \label{subsec:sim-queuing}

As an alternative to his \textit{direct} method --- in this text referred as the constant rate \textit{thinning} method ---, Gillespie introduced the \textit{first reaction} method in his seminal work on simulation algorithms~\cite{gillespie1976}. The \textit{first reaction} method separately simulates the next reaction time for each reaction channel --- \ie~for each mark. It then selects the smallest time as the time of the next event, followed by updating the conditional intensity of all channels accordingly. This is a variation of the constant rate \textit{thinning} method to simulate a set of inter-dependent point processes, making use of the \textit{superposition theorem} --- Theorem 3.3~\cite{last2017} --- in the inverse direction.

Gibson and Bruck~\cite{gibson2000} improved the \textit{first reaction} method with the \textit{next reaction} method. They innovate on three fronts. First, they keep a priority queue to quickly retrieve the next event. Second, they keep a dependency graph to quickly locate all conditional intensity rates that need to be updated after an event is fired. Third, they re-use previously sampled reaction times to update unused reaction times. This minimizes random number generation, which can be costly. Priority queues and dependency graphs have also been used in the context of social media~\cite{farajtabar2017} and epidemics~\cite{holme2021} simulation. In both cases, the phenomena are modelled as point processes.

We prefer to call this class of methods \textit{queuing} methods since most efficiency gains come from maintaining a priority queue of the next event times. Up to this point we assumed that we were sampling from a global process with a mark distribution that could generate any mark \( k \) given an event at time \( t \). With queuing, it is possible to simulate point processes with a finite space of marks as \( M \) interdependent point processes --- see Definition 6.4.1~\cite{daley2003} of multivariate point processes --- doing away with the need to draw from the mark distribution at every event occurrence. Alternatively, it is possible to split the global process into \( M \) interdependent processes each one of which with its own mark distribution.

Our contribution, Algorithm~\ref{algo:sim-queuing}, presents a method for sampling a superposed point process consisting of \( M \) processes by keeping the strike time of each process in a priority queue \( Q \). The priority queue is initially constructed in \( O(M) \) steps in Lines~\ref{line:queuing-init-begin} to~\ref{line:queuing-init-end} of Algorithm~\ref{algo:sim-queuing}. In contrast to \textit{thinning} methods, updates to the conditional intensity depend only on the size of the neighborhood of \( i \). That is, processes \( j \) whose conditional intensity depends on the history of \( i \). If the graph is sparse, then updates will be faster than with \textit{thinning}.

A source of inefficiency in some implementations of queuing algorithms is the fact that one might need to go through multiple rejection cycles before accepting a time candidate \( t_i \) for process \( i \). This might require looking ahead in the future. In addition to that, if process \( j \), which \( i \) depends on, takes place before \( i \), then we need to repeat the whole thinning process to obtain a new time candidate for \( i \). We thus propose Algorithm~\ref{algo:sim-queuing} which is a queuing algorithm that performs thinning in synchrony with the main loop, thus avoiding look ahead and wasted rejections. Since thinning is now synced with the main loop, it is possible to couple this simulator with other algorithms that step chronologically through time. These include ordinary differential equation solvers, enabling us to simulate jump processes with rates given by a differential equation. This is the first synced thinning algorithm we are aware of.

\begin{algorithm}[h]
\begin{algorithmic}[1]
  \Procedure{QueueTime}{\( t \), \( \lambda^\ast \), \( H_{t} \),}
        \State update \( \lambda^\ast \)
        \State find \( \bar{B}^\ast (t) \), \( \ubar{B}^\ast (t) \) and \( L^\ast(t) \) which satisfy Eq.~\ref{eq:thinning-condition}
        \State draw \( u \sim \exp(\bar{B}^\ast(t)) \)
        \If{ \( u > L^\ast(t) \)}
          \State \( \operatorname{accepted} \leftarrow \operatorname{false} \)
        \Else
          \State \( \operatorname{accepted} \leftarrow \operatorname{true} \)
        \EndIf
        \State \( t \leftarrow t + u \)
        \State \Return \( t, \bar{B}^\ast (t), \ubar{B}^\ast, \operatorname{accepted} \)
  \EndProcedure
\end{algorithmic}
\caption{Generates the next candidate time for \textit{queuing}.}
\label{algo:next-time-queuing}
\end{algorithm}

\begin{algorithm}[h]
\begin{algorithmic}[1]
  \Procedure{QueuingMethod}{\( [0, T) \), \( \{ \lambda_{k}^\ast \} \), \( \{ f_{k}^\ast \} \),}
    \State initialize the history \( H_{T^-} \leftarrow \{ \} \)
    \State set \( n \leftarrow 0, t \leftarrow 0 \)
    \For{i=1,M} \label{line:queuing-init-begin}
      \State \( (t_i, \bar{B}_i^\ast, \ubar{B}_i^\ast, a_i) \leftarrow \operatorname{QueueTime}(0, H_{T^-}, \lambda_{i}^\ast(\cdot)) \) \label{line:candidate-one}
      \State push \( (t_i, \bar{B}_i^\ast, \ubar{B}_i^\ast, a_i, i) \) to \( Q \)
    \EndFor \label{line:queuing-init-end}
    \While{\( t < T \)}
      \State first \( (t_i, i, \bar{B}_i^\ast, \ubar{B}_i^\ast, a_i, i) \) from \( Q \)
      \State \( t \leftarrow t_i \)
      \If{\(t \geq T \)}
        \State \textbf{break}
      \EndIf
      \State draw \( v \sim U[0, \bar{B}_i^\ast] \)
      \If{\( ( v > \ubar{B}_i^\ast ) \) and \( ( v > \lambda^\ast \, (t) ) \)}
        \State \( a_i \leftarrow \operatorname{false} \)
      \EndIf
      \If{ \( a_i \)}
        \State \( n \leftarrow n + 1 \)
        \State \( t_n \leftarrow t \)
        \State update  \( f^\ast \) and draw the mark \( k_n \sim f_i^\ast \, (k \mid t_n) \)
        \State update the history \( H_{T^-} \leftarrow H_{T^-} \cup (t_n, k_n) \)
        \For{\( j \in \{ i \} \cup \operatorname{Neighborhood}(i) \)}
          \State \( (t_j, \bar{B}_j^\ast, \ubar{B}_j^\ast, a_j) \leftarrow \operatorname{QueueTime}(0, H_{T^-}, \lambda_{j}^\ast(\cdot)) \)
          \State update \( (t_j, \bar{B}_j^\ast, \ubar{B}_j^\ast, a_j, j) \) in \( Q \)
        \EndFor
      \Else
        \State \( (t_i, \bar{B}_i^\ast, \ubar{B}_i^\ast, a_i) \leftarrow \operatorname{QueueTime}(0, H_{T^-}, \lambda_{i}^\ast(\cdot)) \) \label{line:candidate-one}
 \label{line:candidate-two}
        \State update \( (t_i, \bar{B}_i^\ast, \ubar{B}_i^\ast, a_i, i) \) in \( Q \)
      \EndIf
    \EndWhile
    \State \Return \( H_{T^-} \)
  \EndProcedure
\end{algorithmic}
\caption{The \textit{queuing} method for simulating a marked evolutionary point process over a fixed duration of time \( [0, T) \).}
\label{algo:sim-queuing}
\end{algorithm}

\section{Implementation} \label{sec:implementation}

\texttt{JumpProcesses.jl} is a Julia library for simulating jump --- or point --- processes which is part of Julia's SciML organization.
Our discussion in Section~\ref{sec:method-exact} identified three exact methods for simulating point processes. In all the cases, we identified two mathematical constructs required for simulation: the intensity rate and the mark distribution. In \texttt{JumpProcesses.jl}, these can be mapped to user defined functions \texttt{rate(u, p, t)} and \texttt{affect!(integrator)}. The library provides APIs for defining processes based on the nature of the intensity rate and the intended simulation algorithm. Processes intended for exact methods can choose between \texttt{ConstantRateJump} and \texttt{VariableRateJump}. While the former expects the rate between jumps to be constant, the latter allows for time-dependent rates. The library also provides the \texttt{MassActionJump} API to define large systems of point processes that can be expressed as reaction equations. Finally, \texttt{RegularJump} are intended for inexact methods.

The \textit{inverse} method as described around Equation~\ref{eqn:inverse} uses root find to find the next jump time. Jumps to be simulated via the \textit{inverse} method must be initialized as a \texttt{VariableRateJump}. \texttt{JumpProcesses.jl} builds a continuous callback following the algorithm in~\cite{salis2005} and passes the problem to an \texttt{OrdinaryDiffEq.jl} integrator, which easily interoperates with \texttt{JumpProcesses.jl} (both libraries are part of the \textit{SciML} organization, and by design built to easily compose). \texttt{JumpProcesses.jl} does not yet support the CHV ODE based approach.

Alternatively, \textit{thinning} and \textit{queuing} methods can be simulated via discrete steps. In the context of the library, any method that uses a discrete callback is called an \textit{aggregator}. There are twelve different aggregators, seven of which implement a variation of the \textit{thinning} method and five of which a variation of the \textit{queuing} method.

We start with the \textit{thinning} aggregators, none of which support \texttt{VariableRateJump}. Algorithm~\ref{algo:sim-thinning} assumes that there is a single process. In reality, all the implementations assume a finite multivariate point process with \( M \) interdependent processes. This can be easily conciliated, as we do now, using Definition 6.4.1~\cite{daley2003} which states the equivalence of such process with a point process with a finite space of marks. As all the \textit{thinning} aggregators only deal with \texttt{ConstantRateJump}, the intensity between jumps is constant, Algorithm~\ref{algo:next-time-thinning} short-circuits to quickly return \( t \sim \exp(\bar{B}) = \exp(\lambda_n) \) as discussed in Subsection~\ref{subsec:sim-thinning}. Next, the mark distribution becomes the categorical distribution weighted by the intensity of each process. That is, given an event at time \( t_n \), we have that the probability of drawing process \( i \) out of \( M \) sub-processes is \( \lambda_i^\ast (t_n)  / \lambda^\ast (t_n) \). Conditional on sub-process \( i \), the corresponding \texttt{affect!(integrator)} is invoked, that is, \( k_n \sim f_i^\ast (k \mid t_n) \). Here we use a notation analogous to Section~\ref{subsec:sim-queuing}.

Where most implementations differ is on updating the mark distribution in Line~\ref{line:thinning-mark-sample} of Algorithm~\ref{algo:sim-thinning} and the conditional intensity rate in Line~\ref{line:lambda-update} of Algorithm~\ref{algo:next-time-thinning}. \texttt{Direct} and \texttt{DirectFW} follows the \textit{direct} method in~\cite{gillespie1976} which re-evaluates all intensities after every iteration scaling at \( O(K) \). When drawing the process to fire, it executes a search in an array that stores the cumulative sum of rates. \texttt{DirectCR}, \texttt{SortingDirect} and \texttt{RDirect} only re-evaluate the intensities of the processes that are affected by the realized process. This operation is executed efficiently by keeping a vector of dependencies. These three algorithms differ in how they select the process. \texttt{DirectCR} keeps the intensity rates in a priority table, it is implemented after~\cite{slepoy2008}. \texttt{SortingDirect} keeps the intensity rate in a loosely sorted array following~\cite{mccollum2006}. In both cases, the idea is to use a randomly generated number between zero and one to guide the search for the next jump. With the intensity rates sorted, more frequent processes should be selected faster than less frequent ones. Overall, this should increase the speed of the simulation. \texttt{RDirect} keeps track of the maximum rate of the system, it implements an algorithm equivalent to \textit{thinning} with \( \bar{B} \) equal to the maximum rate. However, the implementation differs. It thins with \( \bar{B} = \lambda_n \), then randomly selects a candidate process and confirms the candidate only if its rate is above a random proportion of the maximum rate. Finally, \texttt{RSSA} and \texttt{RSSACR} group processes with similar rates in bounded brackets. The upper bounds are used for \textit{thinning}. For each round of \textit{thinning}, a sampled candidate process is considered for selection. In \texttt{RSSA}, the candidate process is selected similarly to \texttt{Direct}, while a priority table is used in \texttt{RSSACR}. Both of these algorithms follow from~\cite{thanh2014,thanh2017}.

Next, we consider the \textit{queuing} aggregators. Starting with aggregators that only support \texttt{ConstantRateJump}s we have, \texttt{FRM}, \texttt{FRMFW} and \texttt{NRM}. \texttt{FRM} and \texttt{FRMFW} follow the \textit{first reaction} method in~\cite{gillespie1976}. To compute the next jump, both algorithms compute the time to the next event for each process and select the process with minimum time. This is equivalent to assuming a complete dependency graph in Algorithm~\ref{algo:sim-queuing}. For large systems, they can be less efficient than \texttt{NRM}. The latter implementation is sourced from~\cite{gibson2000} and follows Algorithm~\ref{algo:sim-queuing} very closely.

Previously, we attempted to bridge the gap between the treatment of point process simulation in statistics and biochemistry. Despite the many commonalities, most of the algorithms implemented in \texttt{JumpProcesses.jl} are derived from the biochemistry literature. There has been less emphasis on implementing processes commonly studied in statistics such as self-exciting point processes characterized by time-varying and history-dependent intensity rates. This is addressed by our latest aggregator, \texttt{Coevolve}. This is the first aggregator that supports \texttt{VariableRateJump}s, facilitating substantially more performant simulation of processes with time-dependent intensity rates in \texttt{JumpProcesses.jl} and \texttt{DifferentialEquations.jl} compared to the current \textit{inverse} method-based approach that relies on ODE integration and continuous events.

The implementation of this aggregator takes  inspiration from~\cite{farajtabar2017}, and improves the method in several ways. First, we take advantage of the modularity and composability of Julia to design an API that accepts any intensity rate, not only the Hawkes'. Second, we avoid the re-computation of unused random numbers. When updating processes that have not yet fired, we can transform the unused time of constant rate processes to obtain the next candidate time for the first round of iteration of the \textit{thinning} procedure in Algorithm~\ref{algo:next-time-thinning}. This saves one round of sampling from the exponential distribution, which translates into a faster algorithm. Third, we allow the user to supply a lower bound rate which can short-circuit the loop in Algorithm~\ref{algo:next-time-thinning}, saving yet another round of sampling. Fourth, it adapts to processes with constant intensity between jumps which reduces the loop in Algorithm~\ref{algo:next-time-thinning} to the equivalent implemented in \texttt{NRM}. Finally, since \texttt{Coevolve} can be mapped to a \textit{thinning} algorithm --- see~\cite{farajtabar2017} ---, it can simulate any point process on the real line with a non-negative, left-continuous, history-adapted and locally bounded intensity rate as per Proposition~7.5.I~\cite{daley2003}.

\texttt{Coevolve} syncs with the main execution loop which means that it can be easily coupled with differential equations modeled with \texttt{OrdinaryDiffEq.jl}. For instance, It is possible to model processes whose rates are given by a differential equation. This is a departure from the algorithm described in~\cite{farajtabar2017} which translates not only into a faster, but also more flexible simulator. This difference in implementation follows our discussion on the relationship between the main execution loop and the thinning loop in Section~\ref{subsec:sim-queuing}.

\section{Empirical evaluation} \label{sec:evaluation}

This section conducts some empirical evaluation of the \texttt{JumpProcesses.jl} aggregators described in Section~\ref{sec:implementation}. First, since \texttt{Coevolve} is a new aggregator, we test its correctness by conducting statistical analysis. Second, we conduct the jump benchmarks available in  \texttt{SciMLBenchmarks.jl}. We have added new benchmarks that assess the performance of the new aggregators under settings that could not be simulated with previous aggregators.

\subsection{Statistical analysis of \texttt{Coevolve}}

To simulate a process intended for a discrete solver with \texttt{JumpProcesses.jl}, we define a discrete problem, initialize the jumps and define the jump problem which takes the aggregator as an argument. The jump problem can then be solved with the discrete stepper provided by \texttt{JumpProcesses.jl}, \texttt{SSAStepper}. The code for simulating the homogeneous Poisson process with \texttt{Direct} is reproduced in Listing~\ref{code:sim-poisson}.

\begin{lstlisting}[%
  language = Julia,
  caption = Simulation of the homogeneous Poisson process.,
  label = code:sim-poisson
]
  using JumpProcesses
  rate(u, p, t) = p[1]
  affect!(integrator) = (integrator.u[1] += 1;
    nothing)
  jump = ConstantRateJump(rate, affect!)
  u, tspan, p = [0.], (0., 200.), (0.25,)
  dprob = DiscreteProblem(u, tspan, p)
  jprob = JumpProblem(dprob, Direct(), jump;
    dep_graph=[[1]])
  sol = solve(jprob, SSAStepper())
\end{lstlisting}

The simulation of a Hawkes process --- see Subsection~\ref{subsec:benchmark} for a definition --- requires a \texttt{VariableRateJump} along with the rate bounds and the interval for which the rates are valid. Also, since the Hawkes process is history dependent, we close the \texttt{rate} and \texttt{affect!} function with a vector containing the history of events. The code for simulating the Hawkes process is reproduced in Listing~\ref{code:sim-hawkes}. Note that it is possible to simplify the computation of the rate --- see Subsection~\ref{subsec:benchmark} ---, but we keep the code here as close as possible to its usual definition for illustration purposes.

\begin{lstlisting}[%
  language = Julia,
  caption = Simulation of the Hawkes process.,
  label = code:sim-hawkes
]
  using JumpProcesses
  h = Float64[]
  rate(u, p, t) = p[1] +
    p[2]*sum(exp(-p[3]*(t-_t)) for _t in h; init=0)
  lrate(u, p, t) = p[1]
  urate = rate
  rateinterval(u, p, t) = 1/(2*urate(u,p,t))
  affect!(integrator) = (push!(h, integrator.t);
    integrator.u[1] += 1; nothing)
  jump = VariableRateJump(rate, affect!; lrate,
    urate, rateinterval)
  u, tspan, p = [0.], (0., 200.), (0.25, 0.5, 2.0)
  dprob = DiscreteProblem(u, tspan, p)
  jprob = JumpProblem(dprob, Coevolve(), jump;
    dep_graph=[[1]])
  sol = solve(jprob, SSAStepper())
\end{lstlisting}

To assess the correctness of \texttt{Coevolve}, we add it to the \texttt{JumpProcesses.jl} test suite. Some tests check whether the aggregators are able to obtain empirical statistics close to the expected in a number of simple biochemistry models such as linear reactions, DNA repression, reversible binding and extinction. The test suite was missing a unit test for self-exciting process. Thus, we have added a test for the univariate Hawkes model that checks whether algorithms that accept \texttt{VariableRateJump} are able to produce an empirical distribution of trajectories whose first two moments of the observed rate are close to the expected ones.

\begin{figure}
\subfloat[]{\includegraphics{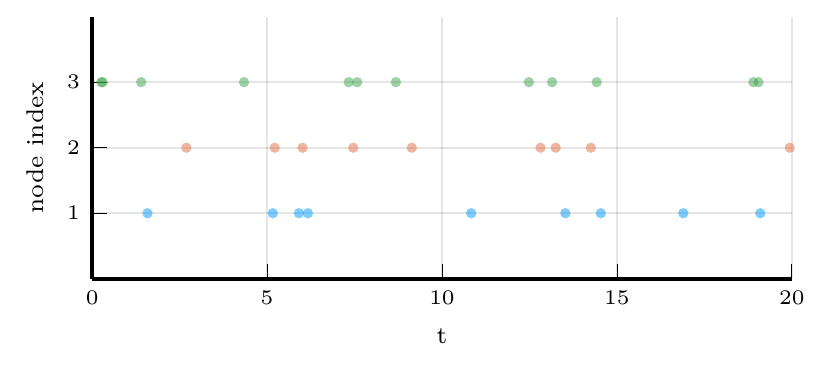}}
\hfil
\subfloat[]{\includegraphics{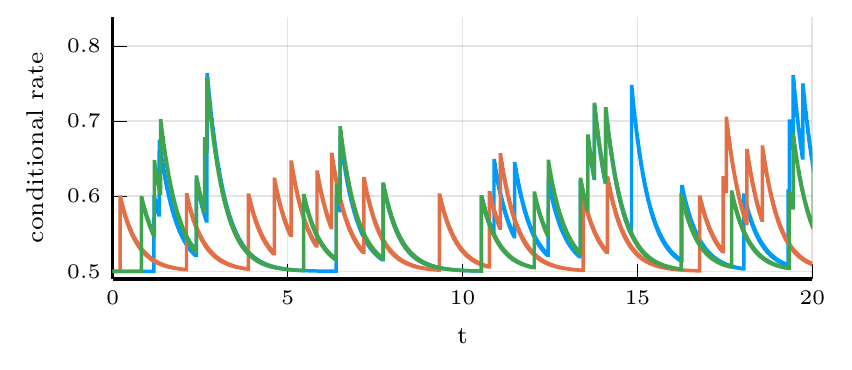}}
\hfil
\subfloat[]{\includegraphics[width=113pt]{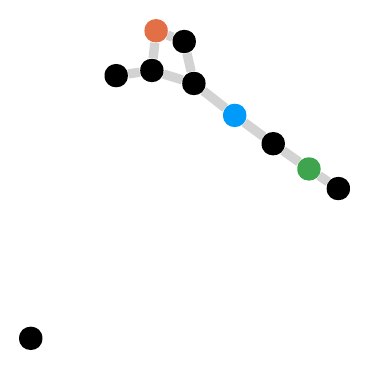}}
\subfloat[]{\includegraphics{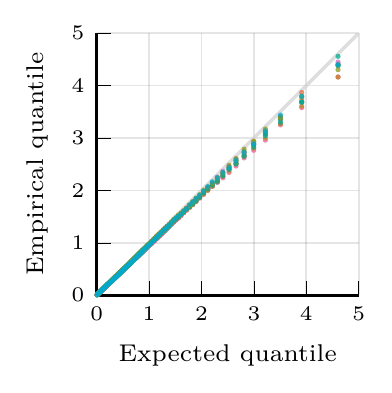}}
\caption{Simulations of 10-nodes compound Hawkes process with parameters \( \lambda = 0.5 , \alpha = 0.1 , \beta = 2.0 \) for \( 200 \) units of time. (a) and (b) sampled trajectory and intensity rate for a single simulation for the three selected nodes in (c) for the first \( 20 \) units of time. (c) underlying 10-nodes network with three random nodes selected. (d) Q-Q plot of transformed inter-event time for 250 simulations colored by node.}
\label{fig:hawkes}
\end{figure}

In addition to that, the correctness of the implemented algorithm can be visually assessed using a Q-Q plot. As discussed in Subsection~\ref{subsec:sim-inverse}, every simple point process can be transformed to a Poisson process with unit rate. This implies that the interval between points for any such transformed process should match the exponential distribution. Therefore, the correctness of any aggregator can be assessed as following. First, transform the simulated intervals with the appropriate compensator. Let \( t_{n_i} \) be the time in which the \( n \)-th event of sub-process \( i \) took place and \( t_{0_i} \equiv 0 \), the compensator for sub-process \( i \) is given by the following:
\begin{equation}
  \Lambda_i^\ast(t_{n_i}) \equiv \Lambda_{n_i}^\ast \equiv \int_0^{t_{n_i}} \lambda_i^\ast(u) du
\end{equation}
Then the transformed simulated interval is given by:
\begin{equation}
  \Delta \Lambda_{n_i} \equiv \Lambda_{n_i}^\ast - \Lambda_{(n-1)_i}^\ast
\end{equation}
Compute the empirical quantiles of the transformed intervals. That is, the \( q \)-th quantile is the interval \( \Delta \Lambda_q \) that divides the sorted intervals in two sets, those below and above \( \Delta \Lambda_q \) such that \( q \)-percent of the elements are below it. Plot the empirical quantiles with the corresponding quantiles of the exponential distribution. If the simulator produces correct trajectories, this plot known as Q-Q plot should depict the points aligned around the 45-degree line. We produce Q-Q plots for the homogeneous Poisson process as well as the compound Hawkes process --- see Subsection~\ref{subsec:benchmark} for a definition --- to attest the correctness of \texttt{Coevolve}. Figure~\ref{fig:hawkes}~(d) depicts the Q-Q plot for a ten-node compound Hawkes process with parameters \( \lambda = 0.5 , \alpha = 0.1 , \beta = 2.0 \) simulated \( 250 \) times for \( 200 \) units of time. Figure~\ref{fig:hawkes} also depicts the trajectory, the conditional intensity and the network structure of a single simulation for three random nodes in panels (a), (b) and (c) respectively. We obtained similar Q-Q plots for the other algorithms that benchmarked the Multivariate Hawkes process below.

\subsection{Benchmarks} \label{subsec:benchmark}

We conduct a set of benchmarks to assess the performance of the \texttt{JumpProcesses.jl} aggregators described in Section~\ref{sec:implementation}. All benchmarks are available in \texttt{SciMLBenchmarks.jl}\footnote{\url{https://github.com/SciML/SciMLBenchmarks.jl/tree/3bf650c1aae7b10e49cbd10e8f626d2a517f3e79/benchmarks/Jumps}}. All were run in BuildKite\footnote{\url{https://buildkite.com/julialang/scimlbenchmarks-dot-jl/builds/1326\#01898353-e5f2-449e-82fd-79708f84462c}} via the continuous integration facilities provided by the package maintainers. We have added two benchmark suites to assess the performance of the new aggregators under settings that could not be simulated with previous aggregators.

First, we assess the speed of the aggregators against jump processes whose rates are constant between jumps. There are four such benchmarks: a 1-dimensional continuous time random walk approximation of a diffusion model (Diffusion), the multi-state model from Appendix A.6~\cite{marchetti2017} (Multi-state), a simple negative feedback gene expression model (Gene I) and the negative feedback gene expression from~\cite{gupta2018} (Gene II). We simulate a single trajectory for each aggregator to visually check that they produce similar trajectories for a given model. The Diffusion, Multi-state, Gene I and Gene II benchmarks are then simulated \( 50 \), \( 100 \), \( 2000 \) and \( 200 \) times, respectively. Check the source code for further implementation details.

Benchmark results are listed in Table~\ref{tab:benchmark-biochemistry}. The table shows that no single aggregator dominates suggesting they should be selected according to the task at hand. However, \texttt{FRM}, \texttt{NRM}, \texttt{Coevolve} never dominate any benchmark. In common, they all belong to the family of queuing methods suggesting that there is a penalty when using such methods for jump processes whose rates are constant between jumps. We also note that the performance of \texttt{Coevolve} lag that of \texttt{NRM} despite the fact that \texttt{Coevolve} should take the same number of steps as \texttt{NRM} when no \texttt{VariableRateJump} is used. The reason behind this discrepancy is likely due to implementation differences, but left for future investigation.

\begin{table}
\centering
\begin{tabular}{lccccc}
\toprule
 & \multicolumn{1}{c}{\textbf{ Diffusion }} & \multicolumn{1}{c}{\textbf{ Multi-state }} & \multicolumn{1}{c}{\textbf{ Gene I }} & \multicolumn{1}{c}{\textbf{ Gene II }} \\
\hline
\textbf{\texttt{Direct}}         & 4.80 s             & 0.11 s             & \textbf{0.17 ms}    & \underline{0.49 s} \\
\textbf{\texttt{FRM}}            & 14.51 s            & 0.20 s             & 0.25 ms             & 0.83 s             \\
\textbf{\texttt{SortingDirect}}  & 1.17 s             & \underline{0.09 s} & \underline{0.18 ms} & \textbf{0.44 s}    \\
\textbf{\texttt{NRM}}            & 0.68 s             & 0.22 s             & 0.33 ms             & 0.82 s             \\
\textbf{\texttt{DirectCR}}       & \underline{0.44 s} & 0.18 s             & 0.35 ms             & 0.87 s             \\
\textbf{\texttt{RSSA}}           & 1.64 s             & \textbf{0.09 s}    & 0.35 ms             & 0.58 s             \\
\textbf{\texttt{RSSACR}}         & \textbf{0.36 s}    & 0.12 s             & 0.72 ms             & 0.91 s             \\
\textbf{\texttt{Coevolve}}       & 0.75 s             & 0.30 s             & 0.45 ms             & 1.13 s             \\
\bottomrule
\end{tabular}
\caption{Median execution time. A 1-dimensional continuous time random walk approximation of a diffusion model (Diffusion), the multi-state model from Appendix A.6~\cite{marchetti2017} (Multi-state), a simple negative feedback gene expression model (Gene I) and the negative feedback gene expression from~\cite{gupta2018} (Gene II). Fastest time is \textbf{bold}, second fastest \underline{underlined}. Benchmark source code and dependencies are available in \texttt{SciMLBenchmarks.jl}, see first paragraph of Section~\ref{subsec:benchmark} for source references.}
\label{tab:benchmark-biochemistry}
\end{table}

Second, we add a new benchmark which simulates the compound Hawkes process for an increasing number processes. Consider a graph with \( V \) nodes. The compound Hawkes process is characterized by \( V \) point processes such that the conditional intensity rate of node \( i \) connected to a set of nodes \( E_i \) in the graph is given by
\begin{equation} \label{eqn:hawkes-brute}
  \lambda_i^\ast (t) = \lambda + \sum_{j \in E_i} \sum_{t_{n_j} < t} \alpha \exp \left[-\beta (t - t_{n_j}) \right].
\end{equation}
This process is known as self-exciting, because the occurrence of an event \( j \) at \( t_{n_j} \) will increase the conditional intensity of all the processes connected to it by \( \alpha \). The excited intensity then decreases at a rate proportional to \( \beta \).
\begin{equation} \label{eqn:hawkes-derivative}
\begin{split}
  \frac{d \lambda_i^\ast (t)}{d t}
    &= -\beta \sum_{j \in E_i} \sum_{t_{n_j} < t} \alpha \exp \left[-\beta (t - t_{n_j}) \right] \\
    &= -\beta \left( \lambda_i^\ast (t) - \lambda \right)
\end{split}
\end{equation}

The conditional intensity of this process has a recursive formulation which can significantly speed the simulation. The recursive formulation for the univariate case is derived in~\cite{laub2021} which also provides additional discussion and results on the Hawkes process. We derive the compound case here. Let \( t_{N_i} = \max \{ t_{n_j} < t \mid j \in E_i \} \) and \( \phi_i^\ast (t) \) below.
\begin{equation}
\begin{split}
  \phi_i^\ast (t)
    % &= \sum_{j \in E_i} \sum_{t_{n_j} < t} \alpha \exp \left[-\beta (t - t_{n_j}) \right] \\
    &= \sum_{j \in E_i} \sum_{t_{n_j} < t} \alpha \exp \left[-\beta (t - t_{N_i} + t_{N_i} - t_{n_j}) \right] \\
    &= \exp \left[ -\beta (t - t_{N_i}) \right] \sum_{j \in E_i} \sum_{t_{n_j} \leq t_{N_i}} \alpha \exp \left[-\beta (t_{N_i} - t_{n_j}) \right] \\
    % &= \exp \left[ -\beta (t - t_{N_i}) \right] \left( \alpha + \sum_{j \in E_i} \sum_{t_{n_j} < t_{N_i}} \alpha \exp \left[-\beta (t_{N_i} - t_{n_j}) \right] \right) \\
    &= \exp \left[ -\beta (t - t_{N_i}) \right] \left( \alpha + \phi_i^\ast (t_{N_i}) \right)
\end{split}
\end{equation}
Then the conditional intensity can be re-written in terms of \( \phi_i^\ast (t_{N_i}) \).
\begin{equation} \label{eqn:hawkes-recursive}
  \lambda_i^\ast (t) = \lambda + \phi_i^\ast (t) = \lambda + \exp \left[ -\beta (t - t_{N_i}) \right] \left( \alpha + \phi_i^\ast (t_{N_i}) \right)
\end{equation}

A random graph is sampled from the Erd\H{o}s-Rényi model. This model assumes the probability of an edge between two nodes is independent of other edges, which we fix at \( 0.2 \). Note that this setup implies an increasing expected node degree.

We fix the Hawkes parameters at \( \lambda = 0.5 , \alpha = 0.1 , \beta = 5.0 \) ensuring the process does not explode and simulate models in the range from \( 1 \) to \( 95 \) nodes for \( 25 \) units of time. We simulate \( 50 \) trajectories with a limit of ten seconds to complete execution. For this benchmark, we save the state of the system exactly after each jump.

We assess the benchmark in eight different settings. First, we run the \textit{inverse} method, \texttt{Coevolve} and \textit{CHV simple} using the brute force formula of the intensity rate which loops through the whole history of past events --- Equation~\ref{eqn:hawkes-brute}. Second, we simulate the same three methods with the recursive formula --- Equation~\ref{eqn:hawkes-recursive}. Next, we run the benchmark against \textit{CHV full}. All \textit{CHV} specifications are implemented with \texttt{PiecewiseDeterministicMarkovProcesses.jl}~\footnote{\url{https://github.com/rveltz/PiecewiseDeterministicMarkovProcesses.jl}} which is developed by Veltz, the author of the \textit{CHV} algorithm discussed in Subsection~\ref{subsec:sim-inverse}. Finally, we run the benchmark using the Python library Tick\footnote{\url{https://github.com/X-DataInitiative/tick}}. This library implements a version of the thinning method for simulating the Hawkes process and implements a recursive algorithm for computing the intensity rate.

Table~\ref{tab:benchmark-hawkes} shows that the \textit{Inverse} method which relies on root finding is the most inefficient of all methods for any system size. For large system size this method is unable to complete all \( 50 \) simulation runs because it needs to find an ever larger number of roots of an ever larger system of differential equations.

The recursive implementation of the intensity rate brings a considerable boost to the simulations, placing \texttt{Coevolve} as one of the fastest algorithms. As shown in Algorithm~\ref{algo:sim-queuing}, every sampled point in \texttt{Coevolve} requires a number of expected updates equal to the expected degree of the dependency graph. Therefore, it is able to complete non-exploding simulations efficiently.

The Python library \texttt{Tick} remains competitive for smaller problems, but gets considerably slower for bigger ones. Also, it is only specialized to the Hawkes process. Another drawback is that the library wraps the actual \texttt{C++} implementation. In contrast, \texttt{JumpProcesses.jl} can simulate many other point processes with a relatively simple user-interface provided by the Julia language.

There is substantial difference between the performance of recursive \textit{CHV simple} and \textit{CHV full}. The former does not make use of the derivative of the intensity function in Equation~\ref{eqn:hawkes-derivative} which is more efficient to compute than the recursive rate in Equation~\ref{eqn:hawkes-recursive}. 

On the one hand, \texttt{Coevolve} clearly dominates \textit{CHV simple}. On the other hand, \textit{CHV full} is slower for smaller networks, but slightly faster than \texttt{Coevolve} for larger models. This change in relative performance occurs due to the rate of rejection in \texttt{Coevolve} increasing in model size for this particular model. We compute the rejection rate as one minus the ratio between the number of jumps and the number of calls to the upper bound. A system with a single node sees a rejection rate of around 8 percent which rapidly increases to 80 percent when the system reaches 20 nodes and plateaus at around 95 percent with 95 nodes.

Finally, we introduce a new benchmark which is intended to assess the performance of algorithms capable of simulating the stochastic model of hippocampal synaptic plasticity with geometrical readout of enzyme dynamics proposed in~\cite{rodrigues2021}. For short, we denote it as the synapse model. We chose to benchmark this model as it is representative of a complex biochemical model. It couples a jump problem containing 98 jumps affecting 49 discrete variables with a stiff, ordinary differential equation problem containing 34 continuous variables. Continuous variables affect jump rates while the discrete variables affect the continuous problem. There are 3 stages to the simulation: pre-synaptic evolution, glutamate release, and post-synaptic evolution. Among the algorithms considered, only the \textit{inverse} method implemented in \texttt{JumpProcesses.jl}, \texttt{Coevolve} and \textit{CHV} are theoretically able to simulate the synapse model. However, in practice, only the last two complete at least one benchmark run. The original synapse problem was described as a piecewise deterministic Markov process, so we do not make the distinction between \textit{CHV simple} and \textit{full} in this benchmark.

Benchmark results are displayed in Table~\ref{tab:benchmark-synapse}. We observe that \textit{CHV} is the fastest algorithm completing the synapse evolution in about half of the time it takes \texttt{Coevolve} with less than half of the allocations. Further investigation reveals that the thinning procedure in \texttt{Coevolve} reaches an average of 70 percent over all jumps which then leads to 2 to 3 times more function evaluations and Jacobians created compared to \textit{CHV}. Our implementation adds stopping times via a call to \texttt{register\_next\_jump\_time!} even for rejected jumps --- we do not know a jump will be rejected until evaluated. This then leads the ODE solver to step to those times and make additional function evaluations and Jacobians. Lemaire~\etal~\cite{lemaire2018} performs a similar benchmark in which they compare the Hodgkin-Huxley model against different thinning conditions and against an ODE approximation. They find that a thinned algorithm with optimal boundary conditions can run significantly faster than the ODE approximation. Thus, there could be plenty of room to improve the performance of \texttt{Coevolve} in our setting.

A disadvantage of \textit{CHV} compared with \texttt{Coevolve} is that it supports limited saving options by design. To save at pre-specified times would require using the fact that solutions are piecewise constant to determine solutions at times in-between jumps --- and for coupled ODE-jump problems would require root-finding to determine when \( s(u) = s_n \) for each desired saving time \( s_n \) in Equation~\ref{eqn:chv-full}. The alternative proposed in~\cite{veltz2015} is to introduce an artificial jump to the model such as the homogeneous Poisson process with unit rate to sample the system at regular intervals. Alternatively, \texttt{Coevolve} allows saving at any arbitrary point. A common workflow in simulating jump processes, particularly when interested in calculating statistics over time, is to pre-specify a precise set of times at which to save a simulation. In theory, this reduces memory pressure, particularly for systems with large numbers of jumps, and can give increased computational performance relative to saving the state at the occurrence of every jump. However, in the case of the synapse model, the number of candidate time rejections far surpasses the number of jumps. Therefore, reducing the number of saving points --- \eg~only saving at start and end of the simulation --- does not significantly reduce allocations or running time. Given these considerations, we decided to save after every jump and at regular pre-specified intervals that occur at the same frequency as the artificial saving jump used by \textit{CHV}.

Another parameter that can affect the precision and speed of the synapse benchmark is the ODE solver. The author of \texttt{PiecewiseDeterministicMarkovProcesses.jl} discuss some of these issues in Discourse\footnote{\url{https://discourse.julialang.org/t/help-me-beat-lsoda/88236}}. Some ODE solvers can be faster and more precise. Due to time constraints, we have not investigated this matter. The synapse benchmark uses the \texttt{AutoTsit5(Rosenbrock23())} solver in both \texttt{Coevolve} and \textit{CHV}. Further investigation of this matter is left to future research.

\begin{table*}
\centering
\begin{tabular}{clllllllll}
\toprule
\multicolumn{1}{l}{} &  & \multicolumn{3}{c}{\textbf{Brute Force }} & \multicolumn{5}{c}{\textbf{Recursive}} \\
\multicolumn{1}{l}{} & \textbf{V} & \textbf{\textit{Inverse}} & \textbf{\texttt{Coevolve}} & \textbf{\textit{CHV}}    & \textbf{\textit{Inverse}} & \textbf{\texttt{Coevolve}} & \textbf{\textit{CHV}}    & \textbf{\textit{CHV}}  & \textbf{\textit{Tick}} \\
\multicolumn{1}{l}{} &            &                           &                            & \textbf{\textit{simple}} &                           &                            & \textbf{\textit{simple}} & \textbf{\textit{full}} &  \\
\hline
\multirow{20}{*}{\textbf{Time}} & \textbf{1}  & 95.9 \( \mu \)s   & \textbf{5.3 \( \bm{\mu} \)s}& 130.3 \( \mu \) s   & 107.9 \( \mu \)s & \underline{6.0 \( \mu \)s}    & 128.6 \( \mu \)s                & 129.4 \( \mu \)s             & 24.7 \( \mu \)s              \\
                                &             &                   &                             &                     &                  &                               &                                 &                              &                              \\
                                & \textbf{10} & 15.0 ms           & 180.6 \( \mu \)s            & 3.8 ms              & 8.2 ms           & \textbf{60.1 \( \bm{\mu} \)s} &  340.6 \( \mu \)s               & 452.8 \( \mu \)s             & \underline{120.2 \( \mu \)s} \\
                                &             &                   &                             &                     &                  &                               &                                 &                              &                              \\
                                & \textbf{20} & 105.2 ms          & 1.5 ms                      & 37.7 ms             & 48.9 ms          & \textbf{223.2 \( \bm{\mu} \)s}&  773.8 \( \mu \)s               & \underline{699.6 \( \mu \)s} & 897.7 \( \mu \)s             \\
                                &             &                   &                             &                     &                  &                               &                                 &                              &                              \\
                                & \textbf{30} & 370.6 ms          & 3.2 ms                      & 101.2 ms            & 155.9 ms         & \textbf{405.8 \( \bm{\mu} \)s}&  1.3 ms                         & \underline{1.1 ms}           & 2.5 ms                       \\
                                &             & \textit{n=28}     &                             &                     &                  &                               &                                 &                              &                              \\
                                & \textbf{40} & 1.7 s             & 7.8 ms                      & 262.2 ms            & 1.1 s            & \textbf{764.4 \( \bm{\mu} \)s}&  2.0 ms                         & \underline{1.4 ms}           & 6.3 ms                       \\
                                &             & \textit{n=7}      &                             & \textit{n=39}       & \textit{n=9}     &                               &                                 &                              &                              \\
                                & \textbf{50} & 3.2 s             & 16.5 ms                     & 556.6 ms            & 2.4 s            & \textbf{1.2 ms}               &  3.0 ms                         & \underline{1.7 ms}           & 13.4 ms                      \\
                                &             & \textit{n=3}      &                             & \textit{n=18}       & \textit{n=5}     &                               &                                 &                              &                              \\
                                & \textbf{60} & 6.3 s             & 32.0 ms                     & 1.0 s               & 4.1 s            & \textbf{1.8 ms}               &  4.3 ms                         & \underline{2.4 ms}           & 27.9 ms                      \\
                                &             & \textit{n=2}      &                             & \textit{n=10}       & \textit{n=3}     &                               &                                 &                              &                              \\
                                & \textbf{70} & 11.5 s            & 52.8 ms                     & 1.8 s               & 6.8 s            & \textbf{2.5 ms}               &  5.7 ms                         & \underline{2.7 ms}           & 56.0 ms                      \\
                                &             & \textit{n=1}      &                             & \textit{n=6}        & \textit{n=2}     &                               &                                 &                              &                              \\
                                & \textbf{80} & 16.6 s            & 88.5 ms                     & 2.8 s               & 11.2 s           & \underline{3.2 ms}            &  7.4 ms                         & \textbf{3.1 ms}              & 93.2 ms                      \\
                                &             & \textit{n=1}      &                             & \textit{n=4}        & \textit{n=1}     &                               &                                 &                              &                              \\
                                & \textbf{90} & 24.9 s            & 124.5 ms                    & 4.8 s               & 15.3 s           & \underline{4.2 ms}            &  9.9 ms                         & \textbf{3.8 ms}              & 152.3 ms                     \\
                                &             & \textit{n=1}      &                             & \textit{n=3}        & \textit{n=1}     &                               &                                 &                              &                              \\
\bottomrule
\end{tabular}
\caption{Median execution time for the compound Hawkes process, V is the number of nodes and n is the total number of successful executions under ten seconds. Brute force refers to the implementation of the intensity rate looping through the whole history of past events. Recursive refers to a recursive implementation that only requires looking at the previous state of each node. \textit{Inverse} and \texttt{Coevolve} are algorithms from \texttt{JumpProcesses.jl}, \textit{CHV} is an algorithm from \texttt{PiecewiseDeterministicMarkovProcesses.jl}. See Subsection~\ref{subsec:sim-inverse} for the distinction between \textit{CHV simple} and \textit{CHV full}. \texttt{Tick} is a Python library. All simulations were run 50 times except when stated otherwise under the running time. Fastest time is \textbf{bold}, second fastest \underline{underlined}. Benchmark source code and dependencies are available in \texttt{SciMLBenchmarks.jl}, see first paragraph of Section~\ref{subsec:benchmark} for source references.}
\label{tab:benchmark-hawkes}
\end{table*}

\begin{table}
\centering
\begin{tabular}{lll}
\toprule
 & \textbf{Time} & \textbf{Allocation}  \\
\hline
\textbf{\textit{Inverse}} & -  & - \\
\textbf{\texttt{Coevolve}} & \underline{3.8 s} & \underline{94.1 Mb}  \\
\textbf{\textit{CHV}} & \textbf{2.0 s} & \textbf{43.5 Mb} \\
\bottomrule
\end{tabular}
\caption{Median execution time and memory allocation. All simulations were run 50 times, a dash indicates that no runs were successful. Fastest time is \textbf{bold}, second fastest \underline{underlined}. Benchmark source code and dependencies are available in \texttt{SciMLBenchmarks.jl}, see first paragraph of Section~\ref{subsec:benchmark} for source references.}
\label{tab:benchmark-synapse}
\end{table}

\section{Conclusion}

This paper demonstrates that \texttt{JumpProcesses.jl} is a fast, general-purpose library for simulating evolutionary point processes. With the addition of \texttt{Coevolve}, any point process on the real line with a non-negative, left-continuous, history-adapted and locally bounded intensity rate can be simulated with this library. The objective of this paper was to bridge the gap between the treatment of point process simulation in statistics and biochemistry. We demonstrated that many of the algorithms developed in biochemistry which served as the basis for the \texttt{JumpProcesses.jl} aggregators can be mapped to three general methods developed in statistics for simulating evolutionary point processes. We showed that the existing aggregators mainly differ in how they update and sample from the intensity rate and mark distribution. As we performed this exercise, we noticed the lack of an efficient aggregator for variable intensity rates in \texttt{JumpProcesses.jl}, a gap which \texttt{Coevolve} is meant to fill.

\texttt{Coevolve} borrows many enhancements from other aggregators in \texttt{JumpProcesses.jl}. However, there are still a number of ways forward. First, given the performance of the \textit{CHV} algorithm in our benchmarks, we should consider adding it to \texttt{JumpProcesses.jl} as another aggregator so that it can benefit from tighter integration with the SciML organization and libraries. The saving behavior of \textit{CHV} might pose a challenge when bringing this algorithm to the library. We could leverage the connection between \textit{inverse} and \textit{thinning} methods illustrated in Subsection~\ref{subsec:sim-thinning} to attempt to develop a version of this algorithm that can evolve in synchrony with model time. Second, the new aggregator depends on the user providing bounds on the jump rates as well as the duration of their validity. In practice, it can be difficult to determine these bounds a priori, particularly for models with many ODE variables. Moreover, determining such bounds from an analytical solution or the underlying ODEs does not guarantee their holding for the numerically computed solution (which is obtained via an ODE discretization), and so modifications may be needed in practice. A possible improvement would be for \texttt{JumpProcesses.jl} to determine these bounds automatically taking into account the derivative of the rates. Deriving efficient bounds require not only knowledge of the problem and a good amount of analytical work, but also knowledge about the numerical integrator. At best, the algorithm can perform significantly slower if a suboptimal bound or interval is used, at worst it can return incorrect results if a bound is incorrect --- \ie~it can be violated inside the calculated interval of validity. Third, \texttt{JumpProcesses.jl} would benefit from further development in inexact methods. At the moment, support is limited to processes with constant rates between jumps and the only solver available \texttt{SimpleTauLeaping} does not support marks. Inexact methods should allow for the simulation of longer periods of time when only an event count per time interval is required. Hawkes processes can be expressed as a branching process. There are simulation algorithms that already take advantage of this structure to leap through time~\cite{laub2021}. It would be important to adapt these algorithms for general, compound branching processes to cater for a larger number of settings. Finally, \texttt{JumpProcesses.jl} also includes algorithms for jumps over two-dimensional spaces. It might be worth conducting a similar comparative exercise to identify algorithms in statistics for \(2 \)- and \( N \)-dimensional processes that could also be added to \texttt{JumpProcess.jl} as it has the potential to become the go-to library for general point process simulation.

\section{Acknowledgements}
This project has been made possible in part by grant number 2021-237457 from the Chan Zuckerberg Initiative DAF, an advised fund of Silicon Valley Community Foundation. SAI was also partially supported by NSF-DMS 1902854.

\bibliographystyle{siam}
\bibliography{references}

\end{document}